\documentclass[12pt]{iopart}

\usepackage{iopams}
\usepackage{setstack}
\usepackage[mathcal]{euscript}
\usepackage{amsgen,amsfonts,amssymb,amsbsy}

\begin{document}

\newcommand{\p}{\partial}
\newcommand{\ri}{\mathrm{i}}
\newcommand{\re}{\mathrm{e}}
\newcommand{\bsf}[1]{\textsf{\textbf{#1}}}
\newcommand{\be}{\begin{equation}}
\newcommand{\ee}{\end{equation}}
\newcommand{\bea}{\begin{eqnarray}}
\newcommand{\eea}{\end{eqnarray}}
\newcommand{\ve}{\varepsilon}
\newcommand{\mC}{\mathcal{C}}
\newcommand{\mm}{\mathbf{m}}

\newcommand{\derivee}[2]{\ensuremath{\frac{\mathrm{d}#1}{\mathrm{d}#2}}}
\newcommand{\deriveep}[2]{\ensuremath{\frac{\partial#1}{\partial#2}}}
\newcommand{\config}[2]{\ensuremath{\mathcal{C}^{#2}_{#1}}}
\newcommand{\average}[2]{\ensuremath{\left\langle {#1} \right\rangle_{#2}}}

\title[]{A mass transport model with a simple non-factorized steady-state distribution}

\author{Jules Guioth and Eric Bertin}

\address{Universit\'e Grenoble Alpes and CNRS, LIPHY, F-38000 Grenoble, France}
\ead{jules.guioth@univ-grenoble-alpes.fr, eric.bertin@univ-grenoble-alpes.fr}

\begin{abstract}
We study a mass transport model on a ring with parallel update, where a continuous mass is randomly redistributed along distinct links of the lattice, choosing at random one of the two partitions at each time step. The redistribution process on a given link depends on the masses on both sites, at variance with the Zero Range Process and its continuous mass generalizations.
We show that the steady-state distribution takes a simple non-factorized form that can be written as a sum of two inhomogeneous product measures.
A factorized measure is recovered for a symmetric mass redistribution, corresponding to an equilibrium process.
A non-equilibrium free energy can be explicitly defined from the partition function.
We evaluate different characterizations of the `distance' to equilibrium, either dynamic or static: the mass flux, the entropy production rate, the Gibbs free-energy difference between the equilibrium and non-equilibrium stationary states, and the derivative of the non-equilibrium free energy with respect to the applied driving force.
The connection between these different non-equilibrium parameters is discussed.
\end{abstract}

\section{Introduction}

One of the goals of non-equilibrium statistical physics is to be able to describe the statistical properties of systems driven in a non-equilibrium steady state by an external non-conservative force.
As no general statistical formalism is available to deal with driven systems,
exactly solvable models have played an important role in the development of this field.
A paradigmatic exactly solvable model is the Asymmetric Simple Exclusion Process (ASEP) \cite{Spitzer}, either with periodic \cite{Derrida98} or open boundary conditions \cite{Derrida,Sandow,Liggett,Schutz}.
Generalizations with several types of particles have also been proposed,
with periodic \cite{Karimipour99,Mallick09,Prolhac12} or open geometries \cite{Vanicat15,Vanicat16,Wheeler16,Vanicat17}.
The ABC model \cite{Mukamel98}, which includes three types of particles, also falls into this class.
The solution of the ASEP model requires in most cases the use of matrix product states \cite{Evans-rev07}, often with infinite size matrices, making its analysis relatively involved. 
Such matrix product state solutions are required even with a periodic geometry, when the model includes several types of particles \cite{Karimipour99,Mallick09,Prolhac12} ---except if some restrictive conditions are imposed \cite{Arita08}.

Simpler models, like the Zero-Range Process (ZRP) \cite{Spitzer,Evans-rev05,Schutz05} and related mass transport models \cite{Evans04,Zia04,Zia06,Bertin06}, have also been considered, often in relation to condensation transitions \cite{Hanney04,Majumdar05,Evans-rev05}. Multispecies generalizations of these models have also been proposed \cite{Hanney04,Hanney06}.
When the transition rates satisfy certain conditions \cite{Evans04,Zia04,Zia06}, these models have the advantage that their steady-state distribution factorizes, making their analytical study much easier.
However, in a closed geometry, they have the drawback that the distribution does not depend on the driving force, and thus remains identical to the equilibrium distribution obtained for unbiased dynamics.
Note that the same property also holds for the (single-species) ASEP on a ring \cite{Derrida98}.

In this paper, we propose a class of mass transport models for which the steady-state distribution takes a simple form (a sum of two inhomogeneous product measures) and explicitly depends on the local driving force.
The present model is inspired by the equilibrium model considered in \cite{Bertin05}, though it differs from the latter in several respects, notably the presence of a driving force and of a synchronous dynamics.
The simple form of the steady-state probability distribution makes calculations easy, as illustrated below on several examples including the evaluation of a non-equilibrium free energy.
In addition, the dependence of the distribution on the forcing allows us to compare dynamical characterizations of the `degree of non-equilibrium' (mass flux and entropy production rate) with static characterizations like the difference of Gibbs free energy functional (or Kullback-Leibler divergence \cite{Khinchin}) between the non-equilibrium distribution and the corresponding equilibrium one.
We also evaluate the non-equilibrium order parameter introduced by Sasa and Tasaki \cite{SasaTasaki06}, defined as a derivative of the non-equilibrium free energy with respect to the driving force, and discuss the relationship between these different measures of the 'distance' to equilibrium.

\section{Definition of the model}
\label{sec:def:model}

We consider a one-dimensional lattice with $N$ sites, labelled by $i=1,\dots,N$, with periodic boundary conditions ($i\pm N \equiv i$); $N$ is assumed to be
even, namely $N=2N'$ with $N'$ integer.
On each site $i$, one defines a real positive mass $m_i$.
The model is endowed with a synchronous, discrete time dynamics\footnote{Note that although the asynchronous, continuous time dynamics is most often used in this context, synchronous dynamics has also been used in the ASEP \cite{Evans97,Rajewsky98,deGier99} and in mass transport models \cite{Evans04,Hanney06}.}
The dynamics proceeds, at each time step $t=0,1,2,\dots$,
by parallel redistributions of mass between neighboring sites $i$ and $i+1$
on one of the two partitions $\mathcal{P}_1 = \{(2k,2k+1)\}$ and $\mathcal{P}_2 = \{(2k+1,2k+2)\}$, randomly chosen with equal probability.
Once a partition $\mathcal{P}_j$ has been selected, all links belonging to the partition $\mathcal{P}_j$ are simultaneously updated.
To update a link $(i,i+1)$, a new value $m_i'$ of the mass on site $i$ is randomly drawn from the distribution
\be \label{eq:def-phi}
\varphi(m_i'|S_i) = \frac{v(m_i') w(S_i-m_i')}{v*w(S_i)},
\qquad S_i \equiv m_i + m_{i+1}
\ee
where $v(m)$ and $w(m)$ are arbitrary positive functions, and $v*w(S)$ is the convolution product of $v$ and $w$,
\be
v*w(S) = \int_0^S \rmd m \, v(m) w(S-m) \,.
\ee
From mass conservation, the mass on site $i+1$ is, after redistribution,
$m_{i+1}' = S_i - m_i'$.

\section{Master equation and steady-state solution}

\subsection{Discrete time master equation}

To describe the statistical evolution of the system under the above dynamics, we write down the corresponding master equation. A configuration of the system is given by the ordered list $\mm=(m_1,\dots,m_N)$ of all the masses in the system. The probability density $P(\mm,t)$ evolves according to the discrete time master equation
\be \label{Master-Equation}
P(\mm',t) = \int \rmd\mm \, T(\mm'|\mm) \, P(\mm,t)
\ee
with $\rmd\mm=\prod_{i=1}^N \rmd m_i$, and
where $T(\mm'|\mm)$ is the probability (density) to jump from configuration $\mm$ to configuration $\mm'$ in a single time step.
This transition probability is normalized according to
\be
\int \rmd\mm' \, T(\mm'|\mm) = 1 \,.
\ee
For the present mass transport model, the transition probability is given by 
\be \label{def:transition}
T(\mm'|\mm) = \frac{1}{2}\, T_1(\mm'|\mm) + \frac{1}{2}\, T_2(\mm'|\mm)
\ee
where
\bea
T_1(\mm'|\mm) &=&
\prod_{k=1}^{N'} \varphi(m_{2k}'|S_{2k}) \,
\delta(S_{2k}' - S_{2k}) \,, \\
T_2(\mm'|\mm) &=&
\prod_{k=1}^{N'} \varphi(m_{2k+1}'|S_{2k+1}) \,
\delta(S_{2k+1}' - S_{2k+1}) \,,
\label{eq:def-T}
\eea
with the shorthand notations
$S_i \equiv m_i + m_{i+1}$ and $S_i' \equiv m_i' + m_{i+1}'$.

\subsection{Steady-state distribution}

In the following, we show that the distribution
\be \label{eq:steady-state-dist}
\fl
P(\mm) = \frac{1}{Z_N(M)} \left( \prod_{k=1}^{N'} v(m_{2k}) w(m_{2k+1})
+ \prod_{k=1}^{N'} w(m_{2k}) v(m_{2k+1}) \right)
\; \delta\left( \sum_{i=1}^N m_i - M \right)
\ee
is a stationary solution of the master equation Eq.~(\ref{Master-Equation}).
In Eq.~(\ref{eq:steady-state-dist}), $M$ is the (constant) total mass,
and $Z_N(M)$ is a normalization factor.
In some cases, it may be convenient to write $P(\mm)$ in the form
$P(\mm) = \frac{1}{2} [ P_1(\mm) + P_2(\mm)]$ with, for $j \in \{1,2\}$
\be \label{eq:def:Pj}
P_j(\mm) = \frac{2}{Z_N(M)} \, Q_j(\mm) \, \delta\left(\sum_{i=1}^N m_i -M\right),
\ee
having defined
\be \label{eq:def:Qj}
Q_1(\mm) = \prod_{k=1}^{N'} v(m_{2k}) w(m_{2k+1}) \,, \qquad
Q_2(\mm) = \prod_{k=1}^{N'} w(m_{2k}) v(m_{2k+1}) \,.
\ee
Using Eq.~(\ref{eq:steady-state-dist}),
the master equation (\ref{Master-Equation}) reads, taking into account the fact that the dynamics conserves the total mass, 
\bea \label{steady-Master-Equation}
&& \!\!\!\!\!\!\!\! Q_1(\mm') + Q_2(\mm') \\ \nonumber
&& \!\! = \frac{1}{2} \int \rmd \mm \, \big[ T_1(\mm'|\mm) 
+ T_2(\mm'|\mm) \big] \, \big[ Q_1(\mm)+Q_2(\mm) \big]
\eea
where, to lighten notations, the Dirac delta function accounting for the total mass conservation is understood.

Expanding the r.h.s.~of Eq.~(\ref{steady-Master-Equation}) into four terms, we evaluate these terms separately, obtaining for $j,k \in \{1,2\}$
(see \ref{app-master-eq})
\be \label{eq:TkQj}
\int \rmd \mm \, T_k(\mm'|\mm) \, Q_j(\mm')
= Q_k(\mm) \,.
\ee
The sum of the four contributions appearing in the r.h.s.~of Eq.~(\ref{steady-Master-Equation}) is thus equal to $Q_1(\mm')+Q_2(\mm')$, so that Eq.~(\ref{steady-Master-Equation}) is satisfied.
Hence the distribution $P(\mm)$ given in Eq.~(\ref{eq:steady-state-dist}) is the stationary solution of the model.

\subsection{Physical interpretation of the dynamics}
\label{sec:phys:dyn}

Without loss of generality, one can rewrite the functions
$v(m)$ and $w(m)$ as
\be \label{param-eps-h}
v(m) = e^{-\beta \ve(m)-\beta h(m)},
\qquad
w(m) = e^{-\beta \ve(m)+\beta h(m)}
\ee
where we have defined
\be
e^{-\beta \ve(m)} = \sqrt{v(m)w(m)},
\qquad
e^{-\beta h(m)} = \sqrt{\frac{v(m)}{w(m)}} \,.
\ee
The parameter $\beta$, to be thought of as an inverse temperature, is arbitrary here, and has only been introduced to facilitate the comparison with equilibrium.
A symmetric redistribution process, obtained for $v(m)=w(m)$,
corresponds to $h(m)=0$, and the stationary distribution Eq.~(\ref{eq:steady-state-dist}) boils down to an equilibrium distribution,
\be \label{eq:equil-dist}
P(\mm) = \frac{2}{Z_N(M)} \, e^{-\beta \sum_{i=1}^N \varepsilon(m_i)}
\; \delta\left( \sum_{i=1}^N m_i -M\right) .
\ee
The function $\ve(m)$ thus appears as an effective local energy associated to a local density $m$.
The function $h(m)$ describes the asymmetry of the dynamics.
In the linear case $h(m) = h_0 m$, having in mind \emph{local detailed balance}, the term $2h_0(m_{i}-m_{i}')$ that enters the ratio $\varphi(m_{i}'|S_{i})/\varphi(m_{i}|S_{i}')$ (with $S_{i}=S_{i}'$) can be interpreted as the work done by a driving force $f=2h_0$ associated with a displaced mass $m_{i}-m_{i}'$ on a unit distance (one lattice spacing).
This case is thus physically meaningful, and we will focus on it when dealing with specific examples (keeping $f$ rather than $h_0$ as the driving parameter).

When $h(m) \ne 0$, the non-equilibrium steady-state distribution $P(\mm)$
given in Eq.~(\ref{steady-Master-Equation}) can be rewritten as
\bea \label{eq:steady-state-dist2}
\fl
P(\mm) = \frac{2}{Z_N(M)} \, e^{-\beta\sum_{i=1}^N \ve(m_i)}
\cosh \left( \sum_{i=1}^N (-1)^i \beta h(m_i)\right)
\; \delta\left( \sum_{i=1}^N m_i - M \right) .
\eea
It may be convenient to rewrite the distribution in a more compact as
\be
P(\mm) = \frac{2}{Z_N(M)} \, e^{-\beta E(\mm)} \, \cosh [\beta H(\mm)]
\; \delta\left( \sum_{i=1}^N m_i - M \right)
\ee
by introducing the global observables
\be \label{eq:def:EH}
E(\mm) = \sum_{i=1}^N \ve(m_i), \qquad H(\mm) = \sum_{i=1}^N (-1)^i h(m_i).
\ee
The presence of the hyperbolic cosine in Eq.~(\ref{eq:steady-state-dist2})
yields long-range correlations, that read to leading order in the driving force
\be
G_j \underset{f\rho\,\ll\,1}{=} \frac{\big(2+(-1)^{j}\big)}{\big(\ve_0-\mu(\rho)\big)^2}\, \rho^2 f^2 + \mathcal{O}\left((\rho f)^4 \right).
\ee
In more intuitive terms, these correlations are generated by the synchronous dynamics over two different partitions of the lattice.
More details on the evaluation of the correlation function and on the expression of the pair and single mass distribution can be found in \ref{app:correl}.

In the following, the arbitrary inverse temperature scale $\beta$ is set to unity, unless stated otherwise.

\section{Non-equilibrium free energy}

\subsection{Large deviation form of the partition function}

It is natural to define from the partition function
\be \label{def-Z}
Z_N(M) = \int \rmd \mm \, [Q_1(\mm) + Q_2(\mm)]
\, \delta\left( \sum_{i=1}^N m_i - M \right)
\ee
a nonequilibrium (intensive) free energy
\be \label{eq:def:Is}
I(\rho) = - \lim_{N \to \infty} \frac{1}{N} \ln Z(N\rho),
\ee
if this limit exists. This means that $Z_N(M)$ takes at large $N$ a large deviation form
\be
Z_N(N\rho) \sim e^{-N I(\rho)}.
\ee
To evaluate $I(\rho)$, we follow the saddle-node method presented
in \cite{Touchette09} (note that the standard G\"artner-Ellis theorem cannot be applied in a straightforward way because $Z_N(M)$ is not a probability distribution). Plugging into Eq.~(\ref{def-Z}) the Laplace representation
of the delta function,
\begin{equation}
  \delta(s) = \frac{1}{2\pi i} \int_{a-i\infty}^{a+i\infty} \!\!\! \rmd \zeta\, e^{\zeta s}
\end{equation}
with $a$ an arbitrary real number, we end up with
\be
Z_N(M) = \frac{1}{2\pi i} \int_{a-i\infty}^{a+i\infty} \!\!\! \rmd \zeta\, e^{-\zeta M}
\left[ \hat{v}(\zeta) \hat{w}(\zeta) \right]^{N/2}
\ee
where
\be
\hat{v}(\zeta) = \int_0^{\infty} \rmd m \, e^{\zeta m} v(m), \qquad
\hat{w}(\zeta) = \int_0^{\infty} \rmd m \, e^{\zeta m} w(m).
\ee
Note that the real part of $\zeta$ (equal to $a$) is chosen small enough for the integrals to converge.
One then has, setting $M=N\rho$ with $\rho$ the average density,
\be \label{eq:Zrho}
Z_N(N\rho) = \frac{1}{2\pi i} \int_{a-i\infty}^{a+i\infty} \!\!\! \rmd\zeta\,
e^{N(\lambda(\zeta)-\rho \zeta)}
\ee
where we have introduced the function\footnote{We take here a determination of the logarithm in the complex plane such that the integration path does not cross the branch cut.}
\be \label{eq:def:lambda}
\lambda(\zeta) = \frac{1}{2}\ln [\hat{v}(\zeta) \hat{w}(\zeta)]
\ee
The function $\lambda(\zeta)$ plays the same role as the scaled cumulant generating function in the G\"artner-Ellis theorem \cite{Touchette09}.
Assuming that $\lambda(\zeta)-\rho \zeta$ has a saddle-point 
$\zeta^{\ast}(\rho)$ defined by
\be \label{eq:zetast:rho}
\frac{d\lambda}{d\zeta}(\zeta^{\ast}) = \rho,
\ee
one can choose $a=\zeta^{\ast}$ in the integral on the rhs of
Eq.~(\ref{eq:Zrho}), leading through a saddle-point evaluation of the integral to
\be
Z_N(N\rho) \sim e^{-N[\rho \zeta^{\ast}(\rho)-\lambda(\zeta^{\ast}(\rho))]}.
\ee
Note that the saddle-point $\zeta^{\ast}$ is necessarily real
because $Z_N$ is real.
The large deviation function $I(\rho)$ introduced
in Eq.~(\ref{eq:def:Is}) is then given by
\be \label{eq:Irho}
I(\rho) = \rho \zeta^{\ast}(\rho)-\lambda(\zeta^{\ast}(\rho)).
\ee
Since the saddle-point $\zeta^{\ast}$ corresponds to a minimum of the function
$\rho \zeta-\lambda(\zeta)$
along a line parallel to the imaginary axis, it also corresponds
to a maximum of this function along the real axis, so that
\be \label{eq:Irho2}
I(\rho) = \sup_{\substack{\zeta \in \mathcal{D}}} \left(\rho\zeta-\lambda(\zeta)\right) . 
\ee
where $\mathcal{D} \subset \mathbb{R}$ is the domain of definition of $\lambda(\zeta)$ over the real axis.

As an example, we evaluate explicitly the large deviation function in the specific case 
of linear functions $\ve(m)=\ve_0 m$ and $h(m)=\frac{1}{2}fm$,
using the parameterization Eq.~(\ref{param-eps-h}) of the functions
$v(m)$ and $w(m)$. One obtains
\be
\hat{v}(\zeta) = \frac{1}{\ve_0 - \zeta + \frac{1}{2}f}, \qquad
\hat{w}(\zeta) = \frac{1}{\ve_0 - \zeta - \frac{1}{2}f}
\ee
and
\be
\lambda(\zeta) = -\frac{1}{2}\ln \left( (\ve_0-\zeta)^{2} - \frac{f^{2}}{4} \right).
\ee
The saddle-point $\zeta^{\ast}(\rho)$, as defined in Eq.(\ref{eq:zetast:rho}), is given by 
\be \label{eq:zetast:expl}
\zeta^{\ast}(\rho) = \ve_0 - \frac{1+\sqrt{1+\rho^{2}f^{2}}}{2\rho},
\ee
so that the large deviation function reads, from Eq.~(\ref{eq:Irho}),
\be \label{eq:Irho:expl}
I(\rho,f) =\ve_0 \rho - \frac{1+\sqrt{1+\rho^{2}f^{2}}}{2} + \frac{1}{2}\ln\left( \frac{1+\sqrt{1+\rho^{2}f^{2}}}{2\rho^{2}}\right) \quad .
\ee
Note that here and in what follows, we emphasize the $f$-dependence of the large deviation function by denoting it as $I(\rho,f)$ when considering the specific case $h(m)=\frac{1}{2}fm$.
At equilibrium, for $f=0$, the large deviation function
reduces to the equilibrium free energy $I(\rho,0) = \epsilon_0 \rho - 1 - \ln\rho$ (we recall that temperature is set to unity).

\subsection{Pressure and chemical potential}

We have seen above that the large deviation function $I(\rho)$ plays the role of a nonequilibrium free energy density.
The associated extensive free energy simply reads
\be \label{eq:def:extens:free-en}
F(M,N) = N\, I\left(\frac{M}{N}\right).
\ee
From this non-equilibrium free energy, one can define, by analogy with equilibrium, a non-equilibrium thermodynamic pressure $p$
and a non-equilibrium chemical potential $\mu$
\cite{SasaTasaki06,Bertin07}
\be
p = - \frac{\partial F}{\partial N}, \qquad \mu = \frac{\partial F}{\partial M}
\ee
(note that $N$ plays here the role of the volume).
From Eq.~(\ref{eq:def:extens:free-en}), one then obtains, using also
Eqs.~(\ref{eq:zetast:rho}) and (\ref{eq:Irho}),
\bea
\mu(\rho) &=& I'(\rho) = \zeta^{\ast}(\rho) \,,\\
p(\rho) &=& -I(\rho) + \rho I'(\rho) \,.
\eea
From these definitions, it follows that the non-equilibrium (intensive) free energy $I(\rho)$ can be expressed as in equilibrium (Euler relation)
\be
I(\rho) = -p(\rho) + \rho\mu(\rho) \,.
\ee
In a non-equilibrium context, this relation was also postulated in
\cite{SasaTasaki06}.
In the example $\ve(m)=\ve_0 m$ and $h(m)=\frac{1}{2}fm$, one finds from
Eq.~(\ref{eq:zetast:expl}) and (\ref{eq:Irho:expl})
\bea
\label{eq:murho:expl}
\mu(\rho) &=& \ve_0 - \frac{1+\sqrt{1+\rho^{2}f^{2}}}{2\rho} \, \\
p(\rho) &=& - \frac{1}{2}\ln\left( \frac{1+\sqrt{1+\rho^{2}f^{2}}}{2\rho^{2}}\right) \,.
\label{eq:prho:expl}
\eea

\section{Dynamical characterization of the `distance' to equilibrium}

We discuss here two different dynamical measures of how far the system is from equilibrium, namely, dynamical quantities that vanish at equilibrium.
Note that the quantities we compute are not necessarily positive, but their absolute value might be interpreted as a `distance' to equilibrium\footnote{We use here the term `distance' in a loose sense, since the quantities considered do not satisfy mathematical properties (like symmetry under exchange) of a true distance.}.
The first quantity is the average mass flux $\Phi$, which is directly related to the bias in the redistribution probability.
The second one is the entropy production rate $\sigma$,
which is rather a measure of the breaking of detailed balance, or in other words, a global measure of probability fluxes in configuration space.

\subsection{Stationary mass flux}

We now evaluate the stationary mass flux between two sites $i$ and $i+1$ (which, due to mass conservation, is independent of $i$).
During a given time step, a mass is transferred between $i$ and $i+1$ only if the link $(i,i+1)$ belongs to the chosen partition ($\mathcal{P}_1$ or $\mathcal{P}_2$) of the lattice; mass transfer thus occurs with probability $\frac{1}{2}$.
The average flux $\Phi$ then reads
\be \label{eq:def:Phi}
\Phi = \frac{1}{2} \big( \langle m_{i} \rangle - \langle m_{i}'\rangle \big)
\ee
where $m_i$ is the mass on site $i$ before a redistribution occurs on the link
$(i,i+1)$, while $m_i'$ is the mass on site $i$ after the redistribution.
The masses $m_i$ and $m_{i+1}$ before redistribution are assumed to follow the steady-state distribution $P(m_i,m_{i+1})$ given in \ref{app:correl} ---see Eq.~(\ref{eq:Pmimi1});
one thus has $\langle m_{i} \rangle =\rho$.
Note that the time step has been set to unity.

The average mass $\langle m_{i}'\rangle$ after redistribution
can be expressed as
\be
\langle m_{i}'\rangle =  \int_0^{\infty} \!\!\! \rmd m_i \int_0^{\infty} \!\!\! \rmd m_{i +1} \,
P(m_i,m_{i+1}) \int_0^{\infty} \!\!\! \rmd m_i' \, m_i' \, \varphi(m_i'|m_i+m_{i+1}).
\ee
After some algebra, one finds
\be
\langle m_{i}'\rangle = 2 C_2(\rho) \int_0^{\infty} \rmd S \, e^{-\mu S}  \int_0^S \rmd m'
\, m' \, v(m') w(S-m') \,.
\ee
The calculation can be carried out explicitly on the example
$\ve(m)=\ve_0 m$ and $h(m)=\frac{1}{2}fm$, yielding
\be
\langle m_{i}'\rangle = \frac{1}{\ve_0-\mu +f}.
\ee
The average mass flux $\Phi$ then reads, using Eqs.~(\ref{eq:murho:expl})
and (\ref{eq:def:Phi}),
\be
\Phi = \frac{1}{2} \big( \rho - \langle m_{i}'\rangle \big)
= \frac{f}{4(\ve_0-\mu)^2 -f^2}.
\ee
Also, using the explicit expression of $\mu(\rho)$ (using Eq.~(\ref{eq:murho:expl})
\begin{equation}
  \label{eq:explicit_exp_flux}
  \Phi = \frac{\rho^{2}f}{2+2\sqrt{1+\rho^{2}f^{2}}} \; .
\end{equation}
Furthermore, one can notice that the flux, which can be interpreted as a response of the system to the driving force $f$ (when $h(m)=\frac{1}{2}fm$), is directly related to the free energy $I$ (\ref{eq:def:Is}) as explicitly shown in \ref{app-flux-free-energy} :
\begin{equation}
  \label{eq:link_flux_free-energy}
  \Phi=-\deriveep{I(\rho,f)}{f}
\end{equation}

\subsection{Entropy production rate}

An alternative dynamical measure of the degree of irreversibility
is given by the entropy production rate.
For a discrete time Markov process,
the (time-dependent) entropy production
rate (i.e., the entropy production per time step) is defined as \cite{Gaspard04}\footnote{Note that notations in \cite{Gaspard04} do not follow the same convention, as $P(\omega|\omega')$ denotes there the probability of a transition from a configuration $\omega$ to a configuration $\omega'$, while we use here a (somehow more standard) conditional probability notation where $T(\mm|\mm')$ is the transition probability from $\mm'$ to $\mm$.}
\be
\fl
\Delta_{\rm int} S_t = \frac{1}{2} \int \rmd \mm \, \rmd \mm' \big[ T(\mm'|\mm) P_t(\mm) -  T(\mm|\mm') P_t(\mm') \big] \ln \frac{T(\mm'|\mm) P_t(\mm)}{T(\mm|\mm') P_t(\mm')} \,.
\ee
The advantage of this form is that the positivity of $\sigma$ is visible, as it involves products of factors of equal sign.
In steady state, the entropy production rate is the opposite of the entropy flow, $\Delta_{\rm int}S = -\Delta_{\rm ext}S$, yielding the simpler expression \cite{Gaspard04}
\be
\Delta_{\rm int} S = \int \rmd \mm\, \rmd \mm' \, T(\mm'|\mm) P(\mm) \ln \frac{T(\mm'|\mm)}{T(\mm|\mm')} \,.
\ee
The entropy production rate $\Delta_{\rm int}S$ can be evaluated in the present model, yielding (technical details are reported in \ref{sec-entropy-prod}):
\be \label{eq:sigma}
\Delta_{\rm int} S = \frac{1}{2} \int \rmd \mm \, \big[ P_1(\mm) - P_2(\mm) \big] \, H(\mm).
\ee
where $P_j(\mm)$ is defined in Eq.~(\ref{eq:def:Pj}).
One thus recovers, as expected, that $\sigma=0$ at equilibrium,
when $P_1(\mm) = P_2(\mm)$.
Eq.~(\ref{eq:sigma}) can be rewritten in terms of the observables
$E$ and $H$ defined in Eq.~(\ref{eq:def:EH}), as
\be
\Delta_{\rm int} S = \frac{1}{Z_N(M)} \int \rmd \mm \, H(\mm)\, e^{-E(\mm)} \, \sinh\big( H(\mm)\big) \, \delta\left( \sum_{i=1}^N m_i -M \right).
\ee
Since the entropy production rate is extensive with system size, it is convenient to define the density $\sigma=\Delta_{\rm int} S/N$ of entropy production rate, in the limit $N \to \infty$.
A way to evaluate $\sigma$ in practice is to introduce the generalized partition function $Z_N(M,\theta)$, obtained by replacing $h(m)$ by $\theta h(m)$ where $\theta$ is a real parameter, yielding
\be
Z_N(M,\theta) = \int \rmd\mm \, e^{-E(\mm)} \, \cosh\big( \theta H(\mm)\big) \, \delta\left( \sum_{i=1}^N m_i -M \right) .
\ee
Assuming a large deviation form $Z_N(N\rho,\theta)\sim e^{-NJ(\rho,\theta)}$, one can then write
\be
\sigma = - \frac{\partial J}{\partial \theta}(\rho,\theta=1) \,.
\ee
The large deviation function $J(\rho,\theta)$ can be evaluated in the same way as $I(\rho)$, simply replacing $h(m)$ by $\theta h(m)$ in the calculation of $\lambda(\zeta)$ ---see Eqs.~(\ref{eq:def:lambda}) and (\ref{eq:Irho2}).

In the specific case $h(m)=\frac{1}{2}fm$, one can also write the entropy production rate in terms of the non-equilibrium free energy $I(\rho,f)$ as
\be
\sigma = -f \frac{\partial I}{\partial f}.
\ee
Given that the flux $\Phi$ is equal to $-\partial I/\partial f$, the entropy production $\sigma$ reads
\be
\sigma = f \, \Phi
\ee
using Eq.~(\ref{eq:link_flux_free-energy}). One then recovers the usual expression of the local entropy production interpreted as the average local work injected in the system (times the inverse temperature that is equal to $1$ here).
Note that if the inverse temperature $\beta \ne 1$, one finds $\sigma = \beta f \, \Phi$.
This result is consistent with the local detailed balance interpretation of the dynamics briefly discussed in Sect.~\ref{sec:phys:dyn}.

\section{Static characterizations of the `distance' to equilibrium}

Having discussed dynamical characterizations of the distance to equilibrium, we turn in this section to static characterizations of this distance, namely, measures of the 'degree of non-equilibrium' that are based only on the steady-state probability distribution $P(\mm)$, without any explicit reference to the dynamics.

\subsection{Difference of Gibbs free energy functional}

One possible such measure is the difference of Gibbs free energy functional between the nonequilibrium and equilibrium distributions, for the same temperature of the thermal bath.
Note that for the sake of clarity, we explicitly take into account in this subsection the temperature $T=\beta^{-1}$ (previously set to $T=1$).
For an arbitrary probability distribution $P(\mm)$ over the configuration space of the model, the Gibbs free energy functional $\mathcal{F}[P]$ is defined as
\be
\mathcal{F}[P] = \int \rmd\mm \, P(\mm) E(\mm) - T \int \rmd \mm \, P(\mm) \ln P(\mm).
\ee
Given that the equilibrium distribution $P_{\rm eq}(\mm)$ at temperature $T$ minimizes the functional $\mathcal{F}[P]$, the quantity
\be \label{eq:def:DeltaF}
\Delta \mathcal{F} = \frac{1}{N} \Big( \mathcal{F}[P] - \mathcal{F}[P_{\rm eq}] \Big)
\ee
satisfies $\Delta \mathcal{F} \ge 0$ for any distribution $P$ (note that we have introduced the factor $1/N$ to make $\Delta \mathcal{F}$ an intensive quantity). It is thus natural to interpret $\Delta \mathcal{F}$ as a measure of the distance to equilibrium. Note that $\beta N \Delta \mathcal{F}$ identifies with the
Kullback-Leibler divergence
\be
\mathbb{D}[P||P_{\rm eq}] = \int \rmd\mm \, P(\mm) \ln \frac{P(\mm)}{P_{\rm eq}(\mm)}.
\ee
In the present model, a straightforward calculation yields
\be \label{eq:DeltaF}
\fl
\Delta \mathcal{F} = \frac{1}{N}\ln Z_N^{\rm eq}(M) -  \frac{1}{N}\ln Z_N(M) 
+ \frac{1}{N} \int \rmd\mm \, P(\mm) \, \ln \cosh [\beta H(\mm)] \,.
\ee
The last integral can be evaluated explicitly in the case $h(m)=\frac{1}{2}fm$,
where one has
\bea \nonumber
&& \int \rmd\mm \, P(\mm) \, \ln \cosh [\beta H(\mm)]\\ \nonumber
&& \qquad \quad = \int \rmd M' \int d\mm \, P(\mm) \, \delta \left( \sum_{k=1}^{N'} m_{2k} -M' \right) \ln \cosh [\beta f(M-2M')]\\
&& \qquad \quad = \int \rmd M' \, \Psi(M'|M) \, \ln \cosh [\beta f(M-2M')]
\label{eq:integ:aux}
\eea
where $\Psi(M'|M)$ is the distribution of the total mass over even sites
$M' = \sum_{k=1}^{N'} m_{2k}$, given the total mass $M$ in the system.
By symmetry, the most probable value of $M'$ is $M/2$, so that by a saddle-point argument, the last integral in Eq.~(\ref{eq:integ:aux}) is equal to zero at order $N$, with only possible subextensive corrections.
One thus finds from Eqs.~(\ref{eq:DeltaF}) and (\ref{eq:def:Is}), for $N \to \infty$,
\be
\Delta \mathcal{F} = I(\rho,f) - I(\rho,0)
\ee
so that $\Delta \mathcal{F}$ also identifies in this case with the difference of free energy as defined by the large deviation function $I(\rho,f)$ of the partition function $Z_N(M)$ ---a quantity a priori distinct from the Gibbs free energy functional, as seen from Eq.~(\ref{eq:DeltaF}).

\subsection{Non-equilibrium order parameter}

A non-equilibrium order parameter $\Psi$ has been introduced by Sasa and Tasaki \cite{SasaTasaki06} as (the opposite of) the derivative of the non-equilibrium free energy with respect to the driving force.
In the present model with $h(m)=\frac{1}{2}fm$, this definition leads to
\be
\Psi = -\frac{\partial I}{\partial f}(\rho,f) \,.
\ee
Several remarks are in order here.
First, this definition is similar to the relation linking, at equilibrium, an order parameter like the magnetization to its conjugate field, hence the name `non-equilibrium order parameter'.
Second, an alternative definition, involving the derivation with respect to the (mass or particle) flux, has also been proposed in \cite{SasaTasaki06}.
Third, we use here an intensive order parameter instead of the extensive order parameter originally introduced in \cite{SasaTasaki06}.

Since the non-equilibrium free-energy $I(\rho,f)$ is, from symmetry arguments, an even function of $f$, $\Psi(\rho,f)$ is an odd function of $f$, and thus vanishes for $f=0$, consistently with the interpretation of $\Psi$ as a non-equilibrium order parameter.

Using Eq.~(\ref{eq:link_flux_free-energy}), the non-equilibrium order parameter $\Psi$ simply boils down to the mass flux,
\begin{equation}
  \label{eq:noneq_order_param}
  \Psi(\rho,f) = \Phi(\rho,f)
\end{equation}
Although $\Psi$ turns out to be equal to $\Phi$, the two quantities differ in essence: $\Psi$ is a static order parameter, while the flux $\Phi$ is a dynamical quantity. Introducing explicitly a time step $\Delta t$ in the model (this time step is been set to $\Delta t=1$ up to now), we would have
$\Phi=\Psi/\Delta t$, showing that both quantities have different dimensions.

\section{Discussion and conclusion}

In this paper, we have introduced a mass transport model with synchronous dynamics for which the steady-state distribution takes a simple non-factorized form, and can be determined explicitly. The knowledge of the steady-state distribution allows for a straightforward evaluation of local distributions of mass, and of a non-equilibrium free energy. The main advantages of this model are on the one hand the simplicity of calculations, and on the other hand the explicit dependence of the steady-state distribution on the driving field ---at odds with, for instance, the ZRP and related mass transport models \cite{Evans-rev05}.

In addition, we have evaluated several quantities, either static or dynamic, that characterize the `degree of non-equilibrium' of the steady state of the system. These include the mass flux $\Phi$, the entropy production rate per site $\sigma$, the difference $\Delta \mathcal{F}$ of Gibbs free energy functional (per site) between the non-equilibrium and equilibrium states,
as well as the non-equilibrium order parameter $\Psi$ introduced by Sasa and Tasaki \cite{SasaTasaki06} as the derivative of the non-equilibrium free-energy with respect to the driving force.
We have found that all these non-equilibrium parameters are closely related one to the other, and that (at least in the case of a density-independent driving force $f$) the non-equilibrium order parameter $\Psi$ may be seen as a key parameter from which the others can be evaluated.
In particular, we have found that
\be
\!\!\!\!\!\!\!\!
\Phi(\rho,f)=\Psi(\rho,f), \qquad \sigma = f \Psi(\rho,f), \qquad
\Delta \mathcal{F}(\rho,f) = \int_0^f \rmd f' \, \Psi(\rho,f') \,.
\ee
For a non-zero applied force $f$, all these parameters have a non-zero value.
The present mass transport model may thus be considered as a genuine non-equilibrium model.
This is to be contrasted, for instance, with more standard mass transport models \cite{Evans04,Evans-rev05} (including the ZRP) which, in spite of the presence of a non-zero particle flux, have vanishing values of $\Psi$ and $\Delta \mathcal{F}$, because their steady-state distribution is independent of the driving.

Future work may consider possible extensions of the model with asynchronous dynamics, where more complicated forms of the steady-state distribution (involving, e.g., matrix-product states) are likely to be needed.
Applications of the model to the field of glassy dynamics could also be considered, by including kinetic constraints in the spirit of the model introduced in \cite{Bertin05}.


\appendix

\section{Evaluation of the integral terms in the master equation}
\label{app-master-eq}

Calculations of the integrals appearing in the steady-state master equation, as formulated in Eq.~(\ref{steady-Master-Equation}), are straightforward.
We provide here the explicit calculation in the case $j=k=1$
[see Eq.~(\ref{eq:TkQj})], using again the short notation
$S_i \equiv m_i + m_{i+1}$ and $S_i' \equiv m_i' + m_{i+1}'$:
\bea
&& \!\!\!\!\!\!\!\!\!\!\!\!\!\!\!\!\!\!
\int \rmd \mm \, T_1(\mm'|\mm) \, Q_1(\mm)
\\ \nonumber
&& \!\!\!\!\!\!\!\!\!\!\!\!\!\!\!\!\!\!
= \frac{1}{Z} \prod_{k=1}^{N'} \int_0^{\infty} \rmd m_{2k} \int_0^{\infty} \rmd m_{2k+1} \, \varphi(m_{2k}'|S_{2k}) \,
v(m_{2k}) w(m_{2k+1}) \,\delta(S_{2k}' - S_{2k})
\\ \nonumber
&& \!\!\!\!\!\!\!\!\!\!\!\!\!\!\!\!\!\!
= \frac{1}{Z} \prod_{k=1}^{N'} \left[ \frac{v(m_{2k}') w(m_{2k+1}')}{v*w(S_{2k}')} \int_0^{\infty} \rmd m_{2k} \int_0^{\infty} \rmd m_{2k+1} \, v(m_{2k}) w(m_{2k+1}) \, \delta(S_{2k}'-S_{2k}) \right].
\eea
Given that
\be \label{eq:app-convol}
\int_0^{\infty} \rmd m_{2k} \int_0^{\infty} \rmd m_{2k+1} \, v(m_{2k}) w(m_{2k+1}) \, \delta(S_{2k}'-S_{2k}) = v*w(S_{2k}')
\ee
one eventually obtains
\be
\int \rmd \mm \, T_1(\mm'|\mm) \, Q_1(\mm)
= Q_1(\mm).
\ee
Calculations for other values of $j,k$ follow the same lines.
For instance, for $k=1$ and $j=2$, $v$ and $w$ are exchanged in the l.h.s.~of
Eq.~(\ref{eq:app-convol}), but the result is the same since the convolution product is commutative.

\section{One- and two-site mass distributions in the thermodynamic limit}
\label{app:correl}

We derive in this appendix the one- and two-site mass distributions
in the limit of an infinitely large system ($N \to \infty$),
keeping the average density $\rho$ fixed.

\subsection{Joint mass distribution on a pair of sites}

The easiest distribution to compute is the joint distribution of masses
$P(m_i,m_{i+1})$ on neighboring sites.
Integrating Eq.~(\ref{eq:steady-state-dist}) over the $N-2$ remaining variables $m_j$ ($j\ne i,i+1$), one finds
\be
\fl
P(m_i,m_{i+1}) = \frac{Z_{N-2}(M-m_i-m_{i+1})}{Z_N(M)} \,
\big[ v(m_i)w(m_{i+1}) + w(m_i)v(m_{i+1}) \big] \,.
\ee
Using the large deviation form of $Z_N$, one finds
\be \label{eq:limit:ZN2:ZN}
\fl
\lim_{N \to \infty} \frac{Z_{N-2}(M-m_i-m_{i+1})}{Z_N(M)} =
\exp\big[ -2I(\rho) + \mu(\rho) ( m_i+m_{i+1}-2\rho)\big] \,.
\ee
Hence the distribution $P(m_i,m_{i+1})$ can be written as
\be \label{eq:Pmimi1}
\fl
P(m_i,m_{i+1}) =  C_2(\rho)\, e^{\mu(\rho) (m_i+m_{i+1})}
\big[ v(m_i)w(m_{i+1}) + w(m_i)v(m_{i+1}) \big] \,,
\ee
where $C_2(\rho)$ is a normalization constant.
It is convenient at this stage to introduce the auxiliary distributions
$p_v(m)$ and $p_w(m)$ defined as
\be
p_v(m) = c_v(\rho)\, e^{\mu(\rho) m} v(m) \,, \qquad p_w(m) = c_w(\rho)\, e^{\mu(\rho) m} w(m),
\ee
where $c_v$ and $c_w$ are normalization constants.
In this way, the distribution $P(m_i,m_{i+1})$ given in Eq.~(\ref{eq:Pmimi1})
can be reformulated as
\be \label{eq:Pmimi1:2}
P(m_i,m_{i+1}) = \frac{1}{2} \big[ p_v(m_i)\,p_w(m_{i+1}) + p_w(m_i)\, p_v(m_{i+1}) \big] \,.
\ee
The same calculation holds for the joint distribution
$P_j(m_i,m_{i+j})$ of the masses $m_i$ and $m_{i+j}$ on distant sites
$i$ and $i+j$, as long as $j$ is odd.
One thus has
\be \label{Pmimij:odd}
\fl
P_j(m_i,m_{i+j}) = \frac{1}{2} \big[ p_v(m_i)\, p_w(m_{i+j}) + p_w(m_i)\, p_v(m_{i+j}) \big] \qquad (j=2k-1,\, k>0) \,.
\ee
When $j$ is even, the calculation is slightly more complicated; one has
\bea
\fl
P_j(m_i,m_{i+j}) = \frac{Z_{N'-2,N'}(M-m_i-m_{i+j})}{Z_N(M)} \, v(m_i)v(m_{i+j}) \\ \nonumber
\qquad \qquad \qquad + \frac{Z_{N',N'-2}(M-m_i-m_{i+j})}{Z_N(M)} \, w(m_i)w(m_{i+j})
\eea
with $N'=N/2$ and where the quantity $Z_{N_1,N_2}(M)$ is defined as
\be
\fl
Z_{N_1,N_2}(M) = \int \prod_{i=1}^{N_1+N_2} \rmd m_i \prod_{i=1}^{N_1} v(m_i)
\prod_{i=N_1+1}^{N_2} w(m_i) \; \delta \left( \sum_{i=1}^{N_1+N_2} m_i -M \right) .
\ee
However, in the limit $N' \to \infty$, the two prefactors
$Z_{N'-2,N'}/Z_N$ and $Z_{N',N'-2}/Z_N$ have the same limit, again given
by Eq.~(\ref{eq:limit:ZN2:ZN}).
Hence the distribution reduces to
\be \label{Pmimij:even}
\fl
P_j(m_i,m_{i+j}) = \frac{1}{2} \big[ p_v(m_i)\, p_v(m_{i+j}) + p_w(m_i)\, p_w(m_{i+j}) \big] \qquad (j=2k,\, k>0) \,.
\ee
Using the more physically meaningful parameterization in terms of the functions
$\ve(\rho)$ and $h(\rho)$, the distribution $P_j(m_i,m_{i+j})$ can also be written for all $j>0$ in the form
\be \label{eq:Pmimi1:3}
\fl
P(m_i,m_{i+1}) =  2 \, C_2(\rho)\, e^{-\ve(m_i)-\ve(m_{i+j})+\mu(\rho) (m_i+m_{i+j})}
\cosh \big[h(m_i)+(-1)^j h(m_{i+j})\big] \,.
\ee
As an explicit example, $P_j(m_i,m_{i+j})$ reads in the specific case
$\ve(m)=\ve_0 m$ and $h(m)=h_0 m$
\bea
\fl
P(m_{i},m_{i+j}) = \frac{\big[ (\ve_0 - \mu(\rho))^{2} - h_0^{2} \big]^{2}}{(\ve_0-\mu(\rho))^{2}+(-1)^{j} h_0^{2}} \, e^{-(\ve_0-\mu(\rho))(m_{i}+m_{i+j})} \\ \nonumber
\qquad \qquad \qquad \qquad \qquad \times \cosh\big( h_0 m_{i} + (-1)^{j}h_0 m_{i+j}\big) 
\eea
where $\mu(\rho)$ is given by Eq.~(\ref{eq:murho:expl}).

\subsection{Two-point correlation}

The two-point correlation function $G_j$ between the masses $m_i$ and $m_{i+j}$,
defined as
\be
G_j
= \langle m_{i}m_{i+j} \rangle - \rho^2
\ee
then takes a simple form.
From Eqs.~(\ref{Pmimij:odd}) and (\ref{Pmimij:even}), one has for $k>0$
\bea
G_{2k-1} &=& \langle m \rangle_v \langle m \rangle_w -\rho^2\\
G_{2k} &=& \frac{1}{2} \big( \langle m \rangle_v ^2 + \langle m \rangle_w^2 \big) -\rho^2
\eea
where $\langle \dots \rangle_v$ and $\langle \dots \rangle_w$ are averages
over the distributions $p_v(m)$ and $p_w(m)$ respectively.
Obviously, $G_j$ is $2$-periodic for $j>0$.
In the example $\ve(m)=\ve_0 m$ and $h(m)=h_0 m$, $G_j$ is given by
\be
G_j = \left( \frac{\rho h_0}{\ve_0 - \mu(\rho)} \right)^2 \frac{h_0^2+ \big(2+(-1)^{j}\big)\big(\ve_0-\mu(\rho)\big)^2}{\big(\ve_0 - \mu(\rho)\big)^2+(-1)^j h_0^2} \, .
\ee
In the limit where $f=2h_0$ is small, one can expand $G_j$ to leading order, yielding
\be
G_j \underset{f\rho\,\ll\,1}{=} \frac{\big(2+(-1)^{j}\big)}{\big(\ve_0-\mu(\rho)\big)^2}\, \rho^2 f^2 + \mathcal{O}\left((\rho f)^4 \right) .
\ee

\subsection{Single-site distribution}

The single-site distribution $p(m)$ is obtained by integrating the two-site distribution over one of the masses. Using for instance Eq.~(\ref{eq:Pmimi1:2}), we get
\be
p(m) = \frac{1}{2} \big[ p_v(m)+p_w(m) \big]
\ee
or equivalently, in terms of $\ve(m)$ and $h(m)$,
\be
p(m) = c(\rho) \, e^{-\ve(m)+\mu(\rho)m} \cosh h(m),
\ee
with $c(\rho)$ a normalization constant.

\section{Link between the flux $\phi$ and the nonequilibrium free energy $I$}
\label{app-flux-free-energy}

When one goes from a configuration $\mm$ to another one $\mm'$, the local instantaneous current $\Delta_{i,i+1}(\mm,\mm')$ that goes to the right on the link $(i,i+1)$ is 
\begin{equation}
  \label{eq:def:instantaneous_local_current}
  \Delta_{i,i+1}(\mm,\mm') = - (m_{i}' - m_{i}) = m_{i+1}'-m_{i+1} \,. 
\end{equation}
Summing over all links, the total mass transfered during the transition $\mm \to \mm'$, $\Delta(\mm,\mm')$, is given by
\begin{equation}
  \label{eq:def:instantaneous_total_mass_transfered}
  \fl
  \Delta(\mm,\mm') = \left\{
  \begin{array}{ll}
   - \sum_{k=1}^{N'} (m_{2k}'-m_{2k})  =   \frac{1}{2} \sum_{i=1}^{N} (-1)^{i}(m_{i}-m_{i}')  & \mathrm{for}\;\mathrm{part.}\; \mathcal{P}_{1} \\
   - \sum_{k=0}^{N'-1} (m_{2k+1}'- m_{2k+1}) =   \frac{1}{2} \sum_{i=1}^{N} (-1)^{i} (m_{i}' - m_{i})  & \mathrm{for}\;\mathrm{part.}\; \mathcal{P}_{2} \\
  \end{array} \right.
\end{equation}
On average,
\begin{eqnarray}
  \label{eq:average_total_mass_transfered}
  \fl
  \average{\Delta(\mm,\mm')}{}  =  \frac{1}{4} \int \!\! \rmd\mm \,\rmd\mm' \left( \sum_{i=1}^{N} (-1)^{i}(m_{i}-m_{i}') \right) T_{1}(\mm'|\mm)P(\mm)  \nonumber \\
                                \qquad \qquad + \frac{1}{4} \int \!\! \rmd\mm \, \rmd\mm' \left( \sum_{i=1}^{N} (-1)^{i}(m_{i}'-m_{i}) \right) T_{2}(\mm'|\mm)P(\mm) \; . 
\end{eqnarray}
Since $\int \rmd\mm' \, T_{k}(\mm'|\mm) = 1$  ($k=1,2$), the terms involving $\sum_{i=1}^{N}m_{i}$ cancel out. Using Eq.~(\ref{eq:def:Qj}) and (\ref{eq:TkQj}), one gets 
\begin{eqnarray}
  \label{eq:average_total_mass_transfered_simplified}
  \average{\Delta(\mm,\mm')}{} =  \frac{1}{2Z_{N}(M)} \int \!\! \rmd\mm' \left( \sum_{i=1}^{N} (-1)^{i}m_{i}' \right) Q_{2}(\mm') \nonumber \\
                                   \qquad \qquad \qquad \qquad - \frac{1}{2Z_{N}(M)} \int \!\! \rmd\mm' \left( \sum_{i=1}^{N} (-1)^{i} m_{i}' \right) Q_{1}(\mm') \; .
\end{eqnarray}
To go further, one needs to use the physical interpretation of the dynamics, given in Sect.~\ref{sec:phys:dyn}. Indeed, using Eq.~(\ref{eq:def:Qj}) and (\ref{param-eps-h}), one can notice that in the (linear) case where $h(m)=\frac{1}{2}fm$, 
\begin{equation}
\fl
  \int \!\! \rmd\mm \left( \sum_{i=1}^{N}(-1)^{i} m_{i} \right) Q_{k}(\mm) =  2 \int \!\! \rmd\mm \frac{H(\mm)}{f} Q_{k}(\mm) =  2 (-1)^{k} \deriveep{Q_{k}}{f}(\mm) \; .
\end{equation}
Eventually, using Eq.~(\ref{def-Z}), the total averaged mass transfered is equal to
\begin{equation}
  \label{eq:final_expression_total_mass_transfered}
  \average{\Delta(\mm,\mm')}{} = \deriveep{\ln Z_{N}}{f} \; ,
\end{equation}
leading to the final expression of the current $\phi$ (mass transfered per site) 
\begin{equation}
  \label{eq:final_expression_current}
  \phi = \frac{\average{\Delta(\mm,\mm')}{}}{N} = \frac{1}{N} \deriveep{\ln Z_{N}}{f} = -\deriveep{I}{f}(\rho,f) \; ,
\end{equation}
thus proving the relation in Eq.~(\ref{eq:link_flux_free-energy}).

\section{Evaluation of the entropy production rate}
\label{sec-entropy-prod}

In this appendix, we evaluate the entropy production in the model defined in
Sect.~\ref{sec:def:model}.
From Eq.~(\ref{def:transition}), the transition rate $T(\mm'|\mm)$ takes the
form
\be
T(\mm'|\mm) = \frac{1}{2} T_1(\mm'|\mm) + \frac{1}{2} T_2(\mm'|\mm)
\ee
where $T_1(\mm'|\mm)$ and $T_2(\mm'|\mm)$ respectively describe redistributions
over the partitions $\mathcal{P}_1$ and $\mathcal{P}_2$ of the lattice. For a given configuration $\mm$, we define the sets
$\mathcal{D}_1(\mm)$ and $\mathcal{D}_2(\mm)$ as the subsets of configurations
$\mm'$ accessible from $\mm$ through redistributions over the partitions $\mathcal{P}_1$ and $\mathcal{P}_2$.
More formally, one has for $j \in \{1,2\}$,
\be
\fl
\mathcal{D}_j(\mm)=\{\mm' | \forall k=1,\dots,N', \; m_{2k+j-1}'+m_{2k+j}'= m_{2k+j-1}+m_{2k+j}\}.
\ee
Using the subsets $\mathcal{D}_1(\mm)$ and $\mathcal{D}_2(\mm)$,
one can express the ratio of reciprocal, nonzero transition probabilities,
so that the entropy production reads, in steady state,
\bea \label{eq:def:sigma:model}
\fl
\Delta_{\rm int} S = \frac{1}{2} \int \rmd \mm \, P(\mm) \left\{ \; \int_{\mathcal{D}_1(\mm)} \rmd \mm'\, T_1(\mm'|\mm) \ln \frac{T_1(\mm'|\mm)}{T_1(\mm|\mm')} \right. \\ \nonumber
 \qquad \qquad \qquad \left. + \int_{\mathcal{D}_2(\mm)} \rmd \mm'\, T_2(\mm'|\mm) \ln \frac{T_2(\mm'|\mm)}{T_2(\mm|\mm')} \right \}.
\eea
The ratios of transition rates can be expressed as
\bea
\ln \frac{T_1(\mm'|\mm)}{T_1(\mm|\mm')} &=& [E(\mm)-E(\mm')]
+[H(\mm)-H(\mm')] \,, \\
\ln \frac{T_2(\mm'|\mm)}{T_2(\mm|\mm')} &=& [E(\mm)-E(\mm')]
-[H(\mm)-H(\mm')] \,.
\eea
The restriction of the integration domains to the subsets
$\mathcal{D}_1(\mm)$ and $\mathcal{D}_2(\mm)$ in Eq.~(\ref{eq:def:sigma:model})
was needed only to be able to properly define the ratio of reverse transition probabilities.
Once Eq.~(\ref{eq:def:sigma:model}) is rewritten in terms of the observables $E(\mm)$ and $H(\mm)$, the integration domains no longer need to be restricted to these subsets since the transition probabilities $T_1(\mm'|\mm)$ and $T_2(\mm'|\mm)$ appearing in the integrals vanish by definition outside the subsets $\mathcal{D}_1(\mm)$ and $\mathcal{D}_2(\mm)$.
Hence one has
\bea \label{eq:def:sigma:model2}
&& \!\!\!\!\!\!\!\!\!\!\!\!\!\!\!\!\!\!\!\!
\Delta_{\rm int} S = \frac{1}{2} \int \rmd \mm \, \rmd\mm' \, P(\mm)
\left[ T_1(\mm'|\mm) \Big( E(\mm)-E(\mm') +H(\mm)-H(\mm') \Big) \right. \\
\nonumber
 && \qquad \qquad \qquad  \left. + T_2(\mm'|\mm) \Big( E(\mm)-E(\mm') -H(\mm)+H(\mm') \Big) \right].
\eea
The part of the integral involving $E$ is easily shown to vanish.
Using the form $P(\mm) = \frac{1}{2} [ P_1(\mm) + P_2(\mm)]$ 
of the probability distribution ---see Eq.~(\ref{eq:def:Pj})---
one has thanks to Eq.~(\ref{eq:TkQj}) that
$\int \rmd \mm \, T_k(\mm'|\mm) P_j(\mm) = P_k(\mm)$.
The $H$-dependent part in Eq.~(\ref{eq:def:sigma:model2}) can then be simplified, after a straightforward calculation, to
\be
\Delta_{\rm int} S = \frac{1}{2} \int \rmd \mm \, \big[ P_1(\mm) - P_2(\mm) \big] \, H(\mm).
\ee
which is precisely Eq.~(\ref{eq:sigma}).

\bigskip

\end{document}